\documentclass[aps,prd,twocolumn,showpacs,superscriptaddress,groupedaddress]{revtex4} 
\usepackage[latin1]{inputenc}
\usepackage{graphicx}
\usepackage{amssymb}
\usepackage{color}
\usepackage{float}
\usepackage{amsmath}
\usepackage{amsfonts}
\usepackage{dcolumn}
\usepackage{hyperref}
\usepackage{amsthm}
\usepackage{color}

\def\3nab{\tilde{\nabla}}

\def\be {\begin{equation}}
\def\ee {\end{equation}}
\def\ba {\begin{eqnarray}}
\def\ea {\end{eqnarray}}

\newcommand{\bra}[1]{\left(#1\right)}

\newcommand{\sfr}[2]{{\textstyle\frac{#1}{#2}}}

\newcommand{\barray}{\begin{array}}
\newcommand{\earray}{\end{array}}

\newcommand{\udot}{{\mathcal A}}

\begin{document}

\title{Accreting Black Holes radiate classical Vaidya radiation to pave way for Hawking radiation}
\author{Naresh Dadhich}
\email{nkd@iucaa.in}
\affiliation{Inter-University Centre for Astronomy \& Astrophysics, Post Bag 4, Pune, 411 007, India}
\affiliation{Astrophysics Research Centre, School of Mathematics, Statistics and Computer Science, University of KwaZulu-Natal, Private Bag X54001, Durban 4000, South Africa}
\author{Rituparno Goswami}
\email{goswami@ukzn.ac.za}
\affiliation{Astrophysics Research Centre, School of Mathematics, Statistics and Computer Science, University of KwaZulu-Natal, Private Bag X54001, Durban 4000, South Africa}
\begin{abstract}
{It is well known that locally defined marginally outer trapped surface (MOTS) is null and coincident with the event horizon of an unperturbed static Schwarzschild black hole. This is however not true for an accreting black hole for which MOTS separates out and turns spacelike. In this letter, we obtain the necessary and sufficient condition for MOTS to remain null and coincident with the event horizon even when matter is continuously accreting on. This also has an important bearing on the quantum Hawking radiation which is supposed to emanate from the MOTS, and it cannot propagate out to infinity unless MOTS is null. The condition is, infalling timelike Type I fluid should turn null or Type II, as it falls on the horizon. This transition from timelike to null is caused by the tidal deformation of the infalling fluid, and that produces an outward directed heat flux giving rise to Vaidya radiation emanating out of the boundary of accreting zone. We thus predict a remarkable new phenomena that accreting black hole radiates classical Vaidya radiation that paves the way for the Hawking radiation. }  

 \end{abstract}


\maketitle

Black holes (BHs) are the most amazing objects in the Universe and they are the quintessential ubiquitous prediction of general relativity (GR). They serve as excellent laboratory for probing the fundamental principles of physics as well as the nature of spacetime and gravity at the deepest level. They are also the most compact objects having the limiting compactness. This is the universal characterisation whether a BH is static or stationary. 

It is well known that BH radiates thermally via the Hawking radiation \cite{Haw74,Haw73,HawPen96,BirDav84}. However the actual site of its emergence, whether from the globally defined null event horizon \cite{ParWil00,Vis01}  or from a a locally defined {\em marginally outer trapped surface} (MOTS), which is the local boundary of the trapped region, still remains a question of intense and engaging debate. In the case of an unperturbed Schwarzschild black hole (which by definition is a global vacuum solution), these surfaces exactly coincide with each other and so there is no problem. However the problem arises when this degeneracy is broken by continuous accretion of the in-falling matter onto the black hole. In this case, if the former is true, that is the source being a globally defined null event horizon, then we encounter a philosophical dilemma for an observer, who needs to wait for infinite proper time to define such a horizon. On the other hand, there is a problem with the latter picture too. Continuous accretion of the in-falling matter (satisfying the physically reasonable energy conditions), onto the black hole makes the MOTS spacelike  \cite{AshKri02}, and no radiation can escape from the points arbitrarily close to these MOTS. There have been numerous attempts to overcome this problem of local boundary of the trapped region. Starting from the idea of the stretched horizon \cite{Suss}, all of these attempts introduced a timelike emitting surface outside the event horizon \cite{His81,Bar14,Hay06}. However, these constructions are quite ad-hoc as it is difficult to motivate the existence of such surfaces geometrically as well physically in an absolutely regular and future asymptotically simple spacetimes. 

So there is a problem, in principle, in our complete understanding of Hawking radiation from an accreting black hole, that stems from the crucial degeneracy breaking caused by in-falling matter.  To resolve this, let's turn the question on its head to ask: \\
{\textbf{Question 1:} Does there exist any important property of spherically symmetric vacuum in GR, that strongly motivates against such degeneracy breaking?} \\
If the answer to the above question is {\em yes}, then the obvious follow up question would be:\\
{\textbf{Question 2:} What does it take to prevent the degeneracy breaking; i.e., to keep both null horizon and MOTS bound together during the accretion process?}\\

{To answer the first question, let us first recall the rigidity of spherical vacuum in GR, as given by  (local) { Birkhoff Theorem}\cite{birk,HawEll73}: {\it Any
$C^2$ solution of Einstein's equations in empty space which is
spherically symmetric in an open set ${\mathcal{S}}$ is locally
equivalent to part of maximally extended Schwarzschild solution in
${\mathcal{S}}$.} In other words any
spherically symmetric solution of the vacuum field equations has
an extra symmetry: it must be either locally static (with a timelike Killing vector (KV)) or spatially
homogeneous (with an extra spacelike KV). This theorem naturally provides a covariant length scale for spherically symmetric vacuum, given by the Schwarzschild mass.}

{In a series of papers \cite{Goswami:2011ft,Goswami:2012jf, Ellis:2013dla}, it was rigorously shown that the rigidity of spherical vacuum solutions of Einstein's
field equations continues even in the perturbed scenario (where the perturbation is small with respect to the covariant scale of the BH mass). That is, for an almost spherically symmetric and almost vacuum spacetimes there always exists an extra almost symmetry with an almost Killing vector that almost solves the Killing equations, where the residual error is much smaller than the covariant scale. If this vector is timelike then the spacetime is locally almost static, and if the vector is spacelike, the spacetime is locally almost spatially homogeneous. This situation is analogous to the one described by Jacobson \cite{Jacob}, where the thermal fluctuations near the Rindler horizon gives rise to approximately flat spacelike 2-surfaces with approximate Killing fields that generates boosts perpendicular to these surfaces and vanishing on them. In cosmological context, this \emph{almost Birkhoff theorem} is in a sense also an analogue of another important result, the \emph{almost EGS theorem} \cite{almEGS}, that generalises from exact isotropy of radiation to approximate isotropy the crucial Ehlers-Geren-Sachs theorem \cite{EGS}. It is obvious then, that for an accreting BH (where the mass of the accreting matter is much smaller than the BH mass), the exterior is almost locally static with a timelike almost KV, whereas the interior is almost locally spatially homogeneous with a spacelike almost KV. Therefore, there must be a null surface that separates these two almost symmetric regions, and that can only be the locally defined MOTS. Thus the existence of these almost symmetries strongly suggests that the MOTS should retain the null character. }

In this letter, we explicitly demonstrate the mechanism by which the degeneracy breaking is avoided, thus answering the second question posed above. We transparently show that when the usual timelike Type I matter, with energy momentum tensor having one timelike and three spacelike eigenvectors \cite{HawEll73}, falls onto the horizon, it must have a phase transition to a null or Type II matter with double null eigenvectors, to avoid the degeneracy breaking. And this transition from Type I to Type II is possible only when the heat flux ($Q$) in the interior of the accreting matter is outgoing and attains the limiting maximum value, which is half of the {\em inertial density} given by sum of the local energy density ($\mu$), isotropic pressure ($p$) and anisotropic stress scalar ($\Pi$) \cite{Dad97, Brassel:2021mje}. We note that it is a limiting value for the heat flux, as any greater value will violate all the energy conditions for the accreting fluid. Interestingly, as if by a magical coincidence, {\it this limiting value of heat flux is exactly the same as required for the MOTS to be null and outgoing \cite{Ellis:2014jja} even when there is matter accreting onto it.} Hence any value other than this will either conflict with the energy conditions or the MOTS-horizon degeneracy. What it means is that the necessary and sufficient condition for the degeneracy of event horizon and MOTS to persist unabated, infalling fluid must turn null as it falls on the horizon. {This coincidence stated above, may stem from a deeper relation between thermodynamics of spacetime and Einstein field equations as described in \cite{Jacob}, where Einstein equations were derived from the proportionality of entropy and the horizon area together with the fundamental relation connecting heat, entropy and temperature. From this perspective, the Einstein equations manifest as equation of state.}

The heat flux, generated by the tidal deformation of accreting fluid, propagates outwards from very close to the horizon through accreting fluid shell which has to be matched at some timelike boundary to the Vaidya radiating metric. Thus we have a remarkable new phenomena that accreting BH must first Vaidya radiate, in order to radiate in the quantum regime. This is indeed very interesting as it is a new prediction of classical radiation from an accreting BH.

In this work we consider spherically symmetric spacetimes, which are a subclass of non-rotating {\em Locally Rotationally Symmetric spacetimes}, that contains continuous non-trivial isotropy group of spatial rotations at every point. Hence, they are equipped with a preferred spatial direction that is covariantly defined at every point. Using the timelike unit vector $u^a$ (defined along the fluid flow lines) and the unit vector along the preferred spatial direction $e^a$ (orthogonal to $u^a$), we can decompose the spacetime as \cite{Ellis:2014jja}
\be\label{decomp}
g_{ab}=-u_au_b+e_ae_b+N_{ab}\,
\ee
where $N_{ab}$ is the 2-dimensional metric that spans the spherical 2-shells. The geometrical quantities defined for the timelike congruence are the expansion scalar $\Theta$, acceleration 3-vector $\dot{u}^a$ and the shear 3-tensor $\sigma_{ab}$. The timelike congruence uniquely defines the electric part of the Weyl tensor, which is responsible for tidal forces and inhomogeneity, as $E_{ab}= C_{acbd}u^cu^d$, while the magnetic part $H_{ab}=C^*_{acbd}u^cu^d$, which {is generated by rotation or by time varying  spacetime }giving rise to gravitational waves, vanishes identically. This timelike congruence $u^a$ can further decompose the energy momentum tensor of the matter to give the energy density $\mu$, isotropic pressure $p$, heat flux 3-vector $q^a$ and anisotropic stress 3-tensor $\pi_{ab}$. The only non-vanishing geometrical quantity related to the preferred spacelike congruence is the volume expansion $\phi$. as the other geometrical quantities related to this congruence vanish identically due to the spherical symmetry. Using this preferred spatial congruence, we can then extract the set of covariant scalars from the above mentioned 3-vectors and 3-tensors  that govern the dynamics of the system as $\mathcal{A}=\dot{u}^ae_a$, $\Sigma=\sigma_{ab}e^ae^b$, $\mathcal{E}=E_{ab}e^ae^b$, $Q=q^ae_a$ and $\Pi=\pi_{ab}e^ae^b$.
Therefore, the set of quantities that fully describe the spherically symmetric class of spacetimes are
\be\label{set1}
\mathcal{D}\equiv\left\{\Theta, \mathcal{A}, \Sigma, \mathcal{E}, \phi, \mu, p, \Pi, Q\right\}\;.
\ee
These geometrical and thermodynamical scalars, together with their directional derivatives along $u^a$ (denoted by a dot) and $e^a$ (denoted by a hat) completely specify the corresponding Ricci identities of $u^a$ and $e^a$ and the Bianchi identities and thereby specify the complete dynamics.  The expression for the Misner-Sharp mass in terms of the covariant scalars is given as
\be
\label{M3}
\mathcal{M} = \frac{1}{2K^{\frac{3}{2}}}\left(\frac{1}{3}\mu - \mathcal{E} - \frac{1}{2}\Pi\right).
\ee
Here $K$ is the Gaussian curvature of the spherical shells. This curvature gives the geometric notion of the area radius of the spherical shells $\mathcal{R}$, as by definition the Gaussian curvature is the inverse square of the area radius. In terms of the covariant scalars, we can write the Gaussian curvature $K$ as  \cite{Ellis:2014jja}
\be
K \equiv\frac{1}{\mathcal{R}^2}=\frac13\mu-\mathcal{E}-\frac12\Pi +\frac14\phi^2
-\bra{\frac13\Theta-\frac12\Sigma}^2\, .\label{gauss}
\ee
From the field equations \cite{Ellis:2014jja}, we can also write down the associated directional derivatives of this mass along $u^a$ and $e^a$ as  
\begin{eqnarray}
\label{MHat}
\hat{\mathcal{M}} &=& \frac{1}{4K^{\frac{3}{2}}}\left[\phi\mu - \left(\Sigma-\frac{2}{3}\Theta\right)Q\right], \\
\label{MDot}
\dot{\mathcal{M}} &=&  \frac{1}{4K^{\frac{3}{2}}}\left[\left(\Sigma - \frac{2}{3}\Theta\right)\left(p + \Pi\right) -\phi Q\right].
\end{eqnarray}
This spacetime decomposition allows us {to} write the energy momentum tensor of a general matter field in the following way
\be\label{EM1}
T_{ab}=\mu u_au_b + (p+\Pi)e_ae_b +2Qu_{(a}e_{b)}+(p-\frac12\Pi)N_{ab}\;.
\ee
In the $[u,e]$ plane (as due to the spherical symmetry the 2-shells are trivial), the discriminant of the quadratic equations for the eigenvalues of the above tensor takes the form
\be\label{eigen1}
\Delta= (\mu+p+\Pi)^2-4Q^2\;.
\ee
{We then }easily see that if $\vert Q\vert < \frac12(\mu+p+\Pi)$, there will be two distinct eigenvalues giving rise to one timelike and one spacelike eigenvectors on this plane. Therefore the matter field is necessarily of Type I or timelike matter field. However, when $\vert Q\vert =\frac12(\mu+p+\Pi)$, then both eigenvalues are equal, which gives double degenerate null eigenvectors and hence the matter field is of Type II or null. Note that the limiting value of the heat flux, $Q$ is fixed by the eigenvalues are double degenerate in 
consonance with the null character of the fluid.

To define MOTS we generalise the definition of dynamical horizons by Ashtekar and
Krishnan \cite{AshKri02} by removing the restriction that it is a
spacelike surface as follows: {\em A  smooth, three-dimensional
sub-manifold $H$ in a spacetime $(\mathcal{M},g)$ is said to be a \textit{Marginally outer trapped surface} if it is foliated by a preferred family of
2-spheres such that, on each leaf $S$, the expansion $\tilde{\Theta}_{out}$ of the outgoing null normal $k^a$ vanishes and the expansion
$\tilde{\Theta}_{in}$ of the other ingoing null normal $l^a$ is strictly negative.}
 
Consequently a MOTS is a 3-manifold, that can be locally timelike, spacelike, or null, which is foliated by marginally trapped 2-spheres. It is shown in \cite{Ellis:2014jja}, that the equation of the MOTS curve in the  local $[u,e]$ plane is given by $\Psi\equiv \bra{\sfr23\Theta-\Sigma+\phi}=0$, and if the vector $\Psi^a=\alpha u^a+\beta e^a$ be the tangent to the curve $\Psi=0$ in the local $[u,e]$ plane, then 
 \be\label{alph}
\frac{\alpha}{\beta}=\frac{\sfr23(\mu+\Lambda)+\sfr12\Pi+{\cal E}-Q}{-\sfr13(\mu+\Lambda)-(p-\Lambda)+{\cal E}-\sfr12 \Pi+Q}\;.
\ee
If ${\alpha}/{\beta}>(<)0$ then the MOTS is said to be future outgoing (ingoing). The nature of the MOTS in terms of it being timelike, spacelike or null can also be determined by $\Psi^a\Psi_a\equiv\beta^2(1-\alpha^2/\beta^2)$. Therefore, ${\alpha^2}/{\beta^2}> 1 (< 1)$ denotes the MOTS to be locally timelike (spacelike). If ${\alpha^2}/{\beta^2}=1$ the MOTS are null.

What happens to the accreting matter in a close vicinity of the MOTS at any given timeslice? In other words, {\it what is the mechanism by which the accreting matter will have an outgoing heat flux?}
 From the expression for the MOTS, it is clear that in this region $\phi\approx (\Sigma-\sfr23\Theta)$. Putting that in equation (\ref{MHat}) we get the expression for the heat flux as
\be\label{Q1}
Q=\mu - \frac{4\hat{\mathcal{M}}}{\phi}K^{\frac{3}{2}}
\ee
Now in the close vicinity of the MOTS we have $\mathcal{R}\approx 2\mathcal{M}$. Therefore the second term of the above equation is $\approx \frac{\hat{\mathcal{M}}}{2\phi\mathcal{M}^3}$. In the realistic case for astrophysical black holes, where the mass of the black hole is much bigger than the change in mass due to accretion, this term must be smaller than the energy density of the accreting matter. This definitely makes the heat flux $Q$ positive and hence {\it outward directed}. Further to this, from the second contracted Bianchi identity we have the following:
\begin{eqnarray}\label{Qdot}
\dot Q &=&-\hat p -\hat\Pi-\bra{\sfr32\phi+\udot}\Pi-\bra{\sfr43\Theta+\Sigma} Q\nonumber\\
 &&   -\bra{\mu+p}\udot\ .
\end{eqnarray}
Again in the physically realistic scenarios the pressures are decreasing away from the MOTS, the expansion scalar for the accreting matter collapsing in a black hole is negative, as well as the acceleration (being directed inward) is negative. Then we can clearly see that the time rate of change of heat flux is positive. {\it This increase in heat flux is the direct consequence of tidal deformations causing the inhomogeneity and anisotropy of the matter, driven by the Weyl curvature, and naturally dictate an outgoing heat flux from the accreting matter which continuously increases as the matter approaches the MOTS.} The heat flux is bounded from {above} when the timelike fluid has attained null character. From Eq. (\ref{eigen1}), the condition for accreting matter turning null at the horizon is $Q=\frac12\left(\mu+p+\Pi\right)$, which then leads, in view of the equation (\ref{alph}), to ${\alpha}/{\beta}=1$ throughout the accretion process and thereby ensuring all time the degeneracy of the MOTS and globally defined event horizon (at the classical level). What will be the consequence for the mass of the black hole inside the spherical horizon? Since on the horizon we have $\left(\Sigma-\frac{2}{3}\Theta\right)=\phi$, substituting the limiting value of the heat flux in equations (\ref{MHat},\ref{MDot}), we get an important result that the 4-gradient of the mass must become null. In other words, we must have $\nabla_a\mathcal{M}\nabla^a\mathcal{M}=-\dot{\mathcal{M}}^2+\hat{\mathcal{M}}^2=0$.

{The outward directed heat flux, that caused the phase transition of the infalling matter at the MOTS, propagates through the accreting zone to its outer boundary. The radius of this outer boundary can be observationally obtained by the cut off frequency of the power law Fourier power spectrum where the long time scale fluctuations are generated in the outer part of the accretion zone \cite{Lyub,Hawkins}. The outer boundary radius can thus be approximated by $R_{\mathcal{B}}\approx\left(\alpha\sqrt{GM}f_{min}\right)^{2/3}$, where $\alpha$ is the viscosity parameter of the accreting fluid and $f_{min}$ is the cut off frequency. For example, in the case of supermassive black hole \emph{ Sagittarius A*}, the horizon radius is approximately $0.08\; AU$, while the radius of the outer boundary of the accretion zone, comprising of cool ionised  gas, was detected to be around $1600\; AU$ \cite{Murchi}.  Beyond the outer boundary, this thermal flux must be radiated away in a pure radiation zone reaching the faraway observer.  Hence, it is natural to match the interior spacetime of accreting matter with an outgoing Vaidya exterior region (that describes a spherically symmetric spacetime with outgoing unpolarised radiation) at this outer boundary of  $R_{\mathcal{B}}$, which must be greater than the area radius of the black hole. }The geometry of the exterior spacetime is described by the Vaidya metric given by
\be
ds_2^2= -\left(1-\frac{2m(v)}{r_v}\right)dv^2-2dvdr_v+r_v^2(d\theta^2+\sin^2(\theta)d\phi^2)\;.
\ee
Here $v$ is the exploding null coordinate, $m(v)$ is the Vaidya mass function and $r_v$ is the Vaidya radius. Geometrically matching the first and the second fundamental forms across the boundary shell of the collapsing accretion ball of area radius $R_{\mathcal{B}}$, we get the following junction conditions
\be\label{match1}
m(v)_{\mathcal{B}}=\mathcal{M}_{\mathcal{B}}\,,
\ee
\be\label{match2}
[m(v)_{,v}]_{\mathcal{B}}= \frac{R^3_{\mathcal{B}}}{R_{\mathcal{B}}-\mathcal{M}_{\mathcal{B}}}\; \left[ \frac12\left(\mu+p+\Pi\right)\right]_{\mathcal{B}}\,,
\ee
where the second condition is the pressure balance condition.

Let us now explain the complete process of spherical accretion onto a black hole step by step to {exhibit} a transparent physical and geometrical perspective of the scenario under consideration:
 
As the spherical shells of accreting matter collapse onto the MOTS, they possess an outward directed heat flux, that attains the limiting maximum value of $Q=\frac12(\mu+p+\Pi)$ in the vicinity of the MOTS. This limiting value of $Q$ is required for fluid to undergo a phase transition, from the timelike to null state at the null MOTS \cite{maeda}.

This limiting value of $Q$, is exactly the value required for the MOTS to be outgoing null in the presence of in-falling matter, just like the unperturbed Schwarzschild black hole in vacuum. Thus, this transition from timelike to null ensures that the situation is exactly the same as it was in the unperturbed case of particle accretion. That is, the degeneracy of local MOTS and the global event horizon {continues} to remain intact.

Moreover, the in-falling energy in the null form, will ensure that the gradient of the mass of the black hole remains null, just as in the unperturbed case. {However, the important difference in the accreting case is that it is achieved at the cost of }the outward directed heat flux. 

The outward directed heat flux from the MOTS will then manifest itself as the classical Vaidya radiation (just like any radiating stellar structure) via a proper spacetime matching on a timelike boundary of the accreting matter zone to an exploding Vaidya region. We note that this Vaidya radiation is not from the MOTS, but from the outer boundary of the accreting zone.
 
{ We note an important point here. Any non-sphericity in the accreting matter (in terms of the various angular momenta) will be treated as the tensor perturbation that evolves according to the background Regge Wheeler potential and will be radiated away via usual quasi normal modes of gravitational waves, {\it generated from the accreting region along with the Vaidya radiation described above. }

We have shown here that the null nature of MOTS in unperturbed Schwarzschild black hole is one of the fundamental features that remains} unaltered in the presence of accreting matter. It is clear that such a scenario is possible within the domain of general relativity, subject to the phase transition of in-falling matter from timelike to null generating heat flux that escapes out as the classical Vaidya radiation from the accretion zone. The phase transition is physically as natural as timelike particle tending to null near the horizon \cite{Dad97,maeda} and dictated by the BH rigidity in terms of the {\it almost Birkhoff theorem} \cite{Goswami:2011ft}.  In this case, the problem of spacelike dynamical horizon goes away, and there is no need to resort to any  fiducial ad-hoc timelike surface as the site of Hawking radiation \cite{Hay06}. The important point is that equivalence of global and local picture of Hawking radiation continues in the perturbed scenario as well. This is how the classical Vaidya {radiation} creates appropriate ground for the quantum Hawking radiation to reach out. The difference between the two radiations, Hawking and Vaidya, besides one being quantum and the other classical,  is the energy radiated out in the former comes from a decrease in BH mass while in the latter it comes from the tidal deformation of fluid.

Several important questions were raised in the context of Hawking radiation for realistic astrophysical black holes in the presence of continuous matter accretion  \cite{Ellis:2014jja}. In the scenario presented above, we are in a position to answer {them all} satisfactorily. The first question about the source of Hawking radiation being local MOTS or global event horizon falls away as the degeneracy of MOTS and global event horizon is never broken. The second question about the source being a timelike or a spacelike surface also goes away, as the null character of Schwarzschild MOTS remains intact even when there is matter accretion.The third question is regarding the reason for the particle creation: the vacuum polarisation or the collapsing accreting matter in a time varying gravitational field as noted long ago \cite{Par69}. It is clear that both the accreting fluid and vacuum polarisation have important roles to play. The thermodynamics of the collapsing fluid generates Vaidya radiation that paves the way {for}  vacuum polarisation to take place that can escape to a far away observer.  
Finally, all this naturally followed from the simple realisation that as fluid approaches the horizon, it turns null thus ensuring the null character of the latter all through the accretion. In the process heat flux is generated in the accreting zone which appears as the Vaidya radiation in the exterior. {This is a new and important effect and {a clear} prediction for an accreting black hole.} 
\begin{acknowledgments} 
We would like to thank the anonymous referee for the invaluable comments, that enormously helped the manuscript to evolve to the present form. ND wishes to thank the University of KwaZulu-Natal (UKZN) for a visit which has facilitated this collaboration. RG is supported by the National Research Foundation (South Africa).
\end{acknowledgments}

 \end{document}